\documentstyle[preprint,prd,eqsecnum,aps,epsf]{revtex}

\tightenlines
\begin{document}

    \newcommand{\DSC}{D\hspace{-0.25cm}\slash_{\bot}}
    \newcommand{\DSP}{D\hspace{-0.25cm}\slash_{\|}}
    \newcommand{\DS}{D\hspace{-0.25cm}\slash}
    \newcommand{\DC}{D_{\bot}}
    \newcommand{\DSCX}{D\hspace{-0.20cm}\slash_{\bot}}
    \newcommand{\DSPX}{D\hspace{-0.20cm}\slash_{\|}}
    \newcommand{\DP}{D_{\|}}
    \newcommand{\QV}{Q_v^{+}}
    \newcommand{\QVB}{\bar{Q}_v^{+}}
    \newcommand{\QVP}{Q^{\prime +}_{v^{\prime}} }
    \newcommand{\QVBP}{\bar{Q}^{\prime +}_{v^{\prime}} }
    \newcommand{\QVHZ}{\hat{Q}^{+}_v}
    \newcommand{\QVHZB}{\bar{\hat{Q}}_v{\vspace{-0.3cm}\hspace{-0.2cm}{^{+}} } }
    \newcommand{\QVPHZB}{\bar{\hat{Q}}_{v^{\prime}}{\vspace{-0.3cm}\hspace{-0.2cm}{^{\prime +}}} }

    \newcommand{\QVPHFB}{\bar{\hat{Q}}_{v^{\prime}}{\vspace{-0.3cm}\hspace{-0.2cm}{^{\prime -}} } }
    \newcommand{\QVPHB}{\bar{\hat{Q}}_{v^{\prime}}{\vspace{-0.3cm}\hspace{-0.2cm}{^{\prime}} }   }

    \newcommand{\QVHF}{\hat{Q}^{-}_v}
    \newcommand{\QVHFB}{\bar{\hat{Q}}_v{\vspace{-0.3cm}\hspace{-0.2cm}{^{-}} }}
    \newcommand{\QVH}{\hat{Q}_v}
    \newcommand{\QVHB}{\bar{\hat{Q}}_v}
    \newcommand{\VS}{v\hspace{-0.2cm}\slash}
    \newcommand{\MQ}{m_{Q}}
    \newcommand{\MQP}{m_{Q^{\prime}}}
    \newcommand{\QVHPMB}{\bar{\hat{Q}}_v{\vspace{-0.3cm}\hspace{-0.2cm}{^{\pm}} }}
    \newcommand{\QVHMPB}{\bar{\hat{Q}}_v{\vspace{-0.3cm}\hspace{-0.2cm}{^{\mp}} }  }
    \newcommand{\QVHPM}{\hat{Q}^{\pm}_v}
    \newcommand{\QVHMP}{\hat{Q}^{\mp}_v}

\newcommand{\PP}{{1 + v\hspace{-0.2cm}\slash \over 2}}
\newcommand{\PM}{{1 - v\hspace{-0.2cm}\slash \over 2}}


\draft
\title{$1/m_Q$ order contributions to $B\to \pi l\nu$ decay in HQEFT}
\author{ W.Y. Wang }
\address{Department of Physics, Tsinghua University, Beijing 100084, China }
 \author{Y.L. Wu  and M. Zhong }
 \address{Institute of Theoretical Physics, Academia Sinica,
 Beijing 100080, China }
\maketitle

\begin{abstract}
Contributions to $B\to \pi l\nu$ decay from $1/m_Q$ order corrections are analyzed in the heavy quark effective
field theory (HQEFT) of QCD. Transition wave functions of $1/m_Q$ order are calculated through light cone sum rule
(LCSR) within the HQEFT framework. The results are compared with those from other approaches.
\end{abstract}


\newpage

\section{Introduction}\label{int}
 Heavy to light semileptonic decay $B\to \pi l\nu$ has attracted much interest due to its special role in
determining the important Cabibbo-Kabayashi-Maskawa (CKM) matrix element $|V_{ub}|$.  A lot of work concerning
this decay may be found in the literatures \cite{var,ar,arsc}. Since $B$ meson consists of a heavy quark and a
light quark, it is natural to explore $B$ decays with considering the heavy quark symmetry. This has been
proved to be powerful in relating various heavy hadron processes, and distinguishing between long and short
distance dynamics reliably.

 The leading order wave functions of the decay $B\to \pi l\nu$ have been calculated in Ref.\cite{bpi} by using light cone
 sum rules within the heavy quark effective field theory (HQEFT)\cite{ylw,wwy,yww,ww,excit}.
In this paper, we present the next to leading order study for $B\to \pi l\nu$ decay in HQEFT. In section
\ref{formulation} we formulate this decay up to $1/m_Q$ order in HQEFT. In section \ref{sumrule} the $1/m_Q$ order
wave functions are calculated by using light cone sum rule method in HQEFT framework. Section \ref{numericalana}
devotes to the numerical analysis and relevant discussions in comparison with other approaches. A short summary is
outlined in section \ref{summary}.

\section{$B\to \pi l\nu$ decay in HQEFT}\label{formulation}

The semileptonic decay $B\to \pi l\nu$ is determined by the matrix
element
\begin{eqnarray}
\label{fdefa}
\langle \pi(p)|\bar{u} \gamma^\mu b|B(p_B) \rangle &=& 2f_{+}(q^2)
p^\mu+(f_{+}(q^2)+f_{-}(q^2))q^\mu  \nonumber\\
&=& f_+(q^2) \Big[p^\mu_B+p^\mu -\frac{m^2_B-m^2_\pi}{q^2}q^\mu \Big]
  +f_0(q^2) \frac{m^2_B-m^2_\pi}{q^2} q^\mu
\end{eqnarray}
with
\begin{eqnarray}
\label{f0ff}
f_0(q^2)=\frac{q^2}{m^2_B-m^2_\pi} f_-(q^2)+f_+(q^2),
\end{eqnarray}
where $q=p_B-p$ is the momentum carried by the lepton pair.

In HQEFT, the heavy quark field $Q$ is decomposed into two parts $Q^+$ and $Q^-$,
 which are formally corresponding to the two solutions of the Dirac equation.
 For the free particle case, they are known to be the quark and antiquark fields
 respectively. In order to shift out the
large component of heavy quark momentum and also to make the leading
order Lagrangian explicitly $m_Q$ independent, an effective heavy quark
field can be conveniently defined as
\newcommand{\VSS}{v\hspace{-0.15cm}\slash}
\begin{equation}
\QV=e^{i\VSS m_Q v\cdot x} \QVHZ =e^{i\VSS m_Q v\cdot x} P_+ Q^{+}
\end{equation}
with $v$ being an arbitrary four-velocity satisfying $v^2=1$, and
$P_+ \equiv (1+\VS)/2$ being a projection operator.
Under this definition, $\QV$ is actually the ``large component"
of heavy quark field. Similarly, one can define the ``large component'' of
heavy antiquark field. This becomes transparent if one considers a free
quark field in the frame $v=(1,0,0,0)$, where one has in the
momentum space
\begin{eqnarray}
\label{component1} \hat{Q}_v^{+} \to \PP u_s(p) &=& \sqrt{E + m \over 2m} \left( \begin{array}{c} 1 \\ 0
\end{array} \right) \varphi_s , \nonumber \\
 \label{component4} \hat{Q}^{-}_v \to \PM v_s(p) &=& \sqrt{E + m \over 2m} \left(
\begin{array}{c} 0 \\ 1 \end{array} \right) \chi_s .
\end{eqnarray}
Here $u_s$ and $v_s$ are the four-component spinors, while $\varphi_s$ and $\chi_s$ are the
two-component Pauli spinor fields that
annihilate a heavy quark and create a heavy antiquark, respectively.
 A detailed discussion on the ``large" and ``small"
components of the heavy quark and antiquark fields can be found in
Refs.\cite{ylw,wwy}.

Integrating out the small component of quark
field as well as the antiquark field,
the effective Lagrangian can be represented in terms of $\QV$ as
\begin{eqnarray}
\label{HQEFTLagrangian}
{\cal L}_{eff}^{(++)}=\QVB iv\cdot D \QV+\frac{1}{m_Q}\QVB(i\DSC)^2\QV
   +{\cal O}(\frac{1}{m^2_Q}) .
\end{eqnarray}
At the same time, a heavy-light quark current
$\bar{q}\Gamma Q$ with $\Gamma$ being arbitrary Dirac matrices
is expanded into
\begin{eqnarray}
\label{HQEFTcE}
\bar{q}\Gamma Q \to e^{-im_Qv\cdot x} \Big\{ \bar{q}\Gamma \QV +\frac{1}{2m_Q}
  \bar{q}\Gamma \frac{1}{i\DSP}(i\DSC)^2 \QV
  +{\cal O}(\frac{1}{m^2_Q}) \Big\},
\end{eqnarray}
where $q$ denotes an arbitrary light quark.
In Eqs.(\ref{HQEFTLagrangian}) and (\ref{HQEFTcE}) the operators
$i\DSP$ and $i\DSC$ are defined by
\begin{eqnarray}
i\DSP=i\VS v\cdot D, \hspace{2cm}
i\DSC=i\DS - i\DSP.
\end{eqnarray}
For detailed derivation of HQEFT,
the interested readers are refered to Refs.\cite{ylw,wwy}.

For the $B\to \pi$ weak transition matrix element, it receives
$1/m_Q$ order contributions not only from the expansion of effective
current (\ref{HQEFTcE}) but also from the insertion of the effective
Lagrangian (\ref{HQEFTLagrangian}).
As a result, the heavy quark expansion (HQE) of the transition
matrix element can be simply written as \cite{wwy}
\begin{eqnarray}
\label{matrixexp}
\langle \pi|\bar{u} \gamma^{\mu} b|B\rangle
&=&\sqrt{\frac{m_B}{\bar{\Lambda}_B}} \Big\{ \langle \pi| \bar{u}\gamma^{\mu}
\QV |M_v \rangle +\frac{1}{2m_Q}\langle \pi|\bar{u}\gamma^{\mu}
\frac{1}{iv\cdot D}P_+ \Big(D^2_{\bot} \nonumber\\
&+&\frac{i}{2}\sigma_{\alpha\beta}
F^{\alpha\beta}\Big) \QV|M_v \rangle +{\cal O}(1/m^2_Q) \Big\} ,
\end{eqnarray}
where $\bar{\Lambda}_B$ is defined by the mass difference of
$B$ meson and bottom quark, $\bar{\Lambda}_B=m_B-m_b$, and
$F^{\alpha\beta}$ is the gluon field strength tensor. $|M_v \rangle$
is an effective heavy meson state defined in HQEFT
to manifest the spin-flavor symmetry,
\begin{eqnarray}
\langle M_v|\QVB \gamma^\mu \QV |M_v \rangle =2\bar{\Lambda}
v^\mu
\end{eqnarray}
with the binding energy $\bar{\Lambda}\equiv
  \lim_{m_Q\to \infty}  \bar{\Lambda}_M$ being heavy flavor
independent.

Note that in Eqs.(\ref{HQEFTcE}) and (\ref{matrixexp})
we have used the operator $1/(iv\cdot D)$ to effectively
represent the contraction
of effective heavy quark fields (or say heavy quark propagator)
\cite{wwy,excit}. In calculating Green functions in the next section, we will
treat this operator as contraction of $\QV$ and $\QVB$.
It is seen in Eq.(\ref{matrixexp}) that the $1/m_Q$ order
corrections to $B\to \pi $ transition are simply attributed to
one kinematic operator and one chromomagnetic operator.

Based on heavy quark spin-flavor symmetry, we parameterize the matrix elements in HQEFT as
\begin{eqnarray}
&&\langle \pi(p) |\bar{u}\Gamma \QV|M_v\rangle =-Tr[\pi(v,p)\Gamma
{\cal M}_v], \\
&&\langle \pi(p) |\bar{u}\Gamma \frac{P_+}{iv\cdot D}D^2_{\bot}\QV
  |M_v\rangle =-Tr[\pi_1(v,p)\Gamma {\cal M}_v], \\
&&\langle \pi(p) |\bar{u}\Gamma \frac{P_+}{iv\cdot D}
   \frac{i}{2}\sigma_{\alpha\beta}F^{\alpha\beta}\QV
  |M_v\rangle =-Tr[\pi^{\alpha\beta}_1(v,p)\Gamma P_+
   \frac{i}{2} \sigma_{\alpha\beta} {\cal M}_v],
\end{eqnarray}
where ${\cal M}_v$ is the flavor independent spin wave function
for pseudoscalar heavy mesons,
\[ {\cal M}_v=-\sqrt{\bar\Lambda} P_+ \gamma^5  .  \]
The functions $\pi(v,p)$, $\pi_1(v,p)$ and
$\pi^{\alpha\beta}_1(v,p)$ can be generally decomposed into
\begin{eqnarray}
\label{parscalar1}
\pi(v,p)&=&\gamma^5 [A(v\cdot p)+ {\hat{p}\hspace{-0.2cm}\slash}
    B(v\cdot p)], \\
\label{parscalar2}
\pi_1(v,p)&=&\gamma^5 [f_a(v\cdot p)+ {\hat{p}\hspace{-0.2cm}\slash}
    f_b(v\cdot p)], \\
\label{parscalar3}
\pi^{\alpha\beta}_1(v,p)&=& \gamma^5 [(\hat{p}^\alpha \gamma^\beta
   -\hat{p}^\beta \gamma^\alpha)(s_1(v\cdot p)+{\hat{p}\hspace{-0.2cm}\slash} s_2(v\cdot p))
  +i\sigma^{\alpha\beta} (s_3(v\cdot p)+{\hat{p}\hspace{-0.2cm}\slash} s_4(v\cdot p))]
\end{eqnarray}
with $\hat{p}^\mu =p^\mu /( v\cdot p)$.
The Lorentz scalar functions $A$, $B$, $f_a$, $f_b$
and $s_i(i=1,2,3,4)$
are independent of the heavy quark mass.
Their scale dependence is suppressed in these formulae.
In Eqs.(\ref{parscalar1})-(\ref{parscalar3})
we use for convenience $v\cdot p$ as the variable
of wave functions, which is related to the momentum transfer
$q^2$ by
\[ \xi  \equiv v\cdot p =\frac{m^2_B+m^2_\pi-q^2}{2m_B}  . \]

Now after trace calculation Eq.(\ref{matrixexp}) yields
\begin{eqnarray}
\label{matrixinfac}
\langle \pi(p)|\bar{u}\gamma^\mu b|B(p_B)\rangle
 =2 \sqrt{\frac{m_B \bar\Lambda}{\bar{\Lambda}_B}} \Big\{ v^\mu
 \Big(A+\frac{1}{2m_Q}A_1 \Big)+
  \hat{p}^{\mu} \Big(B+\frac{1}{2m_Q}B_1 \Big)
 +{\cal O}\Big(\frac{1}{m^2_Q} \Big) \Big\}
\end{eqnarray}
with
\begin{eqnarray}
A_1&=&f_a+2s_1-2s_2 \frac{m^2_\pi}{ \xi^2} -3s_3, \\
B_1&=&f_b-2s_1+2s_2-3s_4 ,
\end{eqnarray}
where $p^2=m^2_\pi$ is used.
Eqs.(\ref{fdefa}) and (\ref{matrixinfac}) yield relations between
the form factors $f_{\pm}$ and the universal wave functions,
\begin{eqnarray}
\label{AandL1}
 f_{\pm}(q^2)&=& \sqrt{ \frac{\bar\Lambda}
 {m_B \bar{\Lambda}_B } }
 \Big\{ \Big[A(\xi)+\frac{1}{2m_Q}A_1(\xi) \Big] \pm
   \Big[B(\xi)+\frac{1}{2m_Q}B_1(\xi ) \Big] \frac{m_B}{\xi }
   +{\cal O}\Big(\frac{1}{m^2_Q}\Big) \Big\} .
\end{eqnarray}

 As a comparison, in the usual heavy quark effective theory (HQET) one uses the effective Lagrangian
\begin{eqnarray}
\label{HQETLagr}
{\cal L}_{eff}|_{HQET}=\QVB iv\cdot D \QV+\frac{1}{2m_Q}\QVB (i\DSC)^2
  \QV +{\cal O} \Big(\frac{1}{m^2_Q} \Big) ,
\end{eqnarray}
and the heavy-light current expansion (at tree level)
\begin{equation}
\label{HQETcE}
\bar{q}\Gamma Q^+\to \bar{q}\Gamma \QV+\frac{1}{2m_Q}\bar{q}
  \Gamma i\DSC \QV +{\cal O} \Big(\frac{1}{m^2_Q} \Big) .
\end{equation}
In some references the notation $h_v$ is used for the field
variable $\QV$ here.

For the heavy quark expansion of the transition matrix element in HQET, the $1/m_Q$ order
corrections from the current expansion (\ref{HQETcE}) and those
from the insertion of the Lagrangian (\ref{HQETLagr}) can not be
attributed to the same set of operators. In particular, there are both
operators with even and odd powers of $i\DSC$ which contribute at $1/m_Q$ order.
Therefore the corrections from these
two sources have to be characterized by different sets of wave
functions and considered separately.

In HQEFT, however, there are only
operators with even power of $i\DSC$ appearing in the effective
Lagrangian (\ref{HQEFTLagrangian}) and current (\ref{HQEFTcE}). The
terms with odd powers of \mbox{$i\DSC$} are canceled due to the
inclusion of the antiquark contributions.
Therefore the $1/m_Q$ order corrections to transition matrix elements
from the two sources
can be attributed to the same set of operators,
and the final formulation in HQEFT is simple \cite{wwy}.
In this paper, we need only
calculate the two composite functions $A_1$ and $B_1$.

\section{Light cone sum rules in HQEFT}\label{sumrule}

To calculate the $1/m_Q$ order wave functions $A_1$ and $B_1$, we consider
the two-point correlation function
\begin{eqnarray}
\label{correlator}
F^\mu&=&i\int d^4x e^{i(q-m_Qv)\cdot x}\langle \pi(p)|T\Big\{\bar{u}(x)\gamma^\mu
 \frac{1}{iv\cdot D}P_+ \Big(D^2_\bot +\frac{i}{2} \sigma_{\alpha\beta}
 F^{\alpha\beta}\Big)\QV(x), \nonumber\\
&& \QVB(0)i\gamma^5 d(0) \Big\}|0 \rangle
\end{eqnarray}
where $\QVB(0) i\gamma^5 d(0)$
is the interpolating current for $B$ meson.
Inserting between the two currents in Eq.(\ref{correlator}) a complete
set of intermediate states with the $B$ meson quantum number,
one gets
\begin{eqnarray}
\label{correlatorhad}
 2iF
 \frac{A_1 v^\mu+B_1 \hat{p}^\mu}{2\bar{\Lambda}_B- \omega}
 +\int^\infty_{s_0} ds \frac{ \rho_a(v\cdot p,s)
 v^\mu+\rho_b(v\cdot p,s) \hat{p}^\mu } {s- \omega}
 +subtraction
\end{eqnarray}
with $\omega \equiv 2v\cdot k$, where $k=p_B-m_Q v$ is the residual
momentum. $F$ is the decay constant of $B$ meson at the leading order
of $1/m_Q$ expansion, defined by \cite{ww}
\begin{equation}
\langle 0|\bar{q}\Gamma \QV|B_v \rangle =\frac{F}{2}Tr[\Gamma {\cal M}_v].
\end{equation}
At the same time, in deep Euclidean region the correlator
(\ref{correlator}) can be
calculated in effective field theory. The result can
be written as an integral over a theoretic spectral
density,
\begin{equation}
\label{correlatorth}
\int^{\infty}_{0} ds
\frac{\rho^{th}_a( \xi,s) v^\mu+\rho^{th}_b(\xi ,s) \hat{p}^\mu}
{s- \omega}+subtraction .
\end{equation}

In the sum rule analysis, one assumes the quark-hadron duality,
and equals the hadronic representation (\ref{correlatorhad}) and the
theoretic one (\ref{correlatorth}), which provides a sum rule
equation. Furthermore, in order to enhance the importance of
the ground state contribution, to suppress higher order
nonperturbative contributions and also to remove the $subtraction$,
a Borel transformation
\[
\hat{B}^{(\omega)}_{T}\equiv \lim_{
{\tiny \begin{array}{c}
-\omega,n\to \infty \\
  -\omega/n=T  \end{array} }  }
 \frac{(-\omega)^{n+1}}{n!}(\frac{d}{d\omega})^n
\]
should be performed to both
sides of the sum rule equation.
With using the formulae
\begin{eqnarray}
\label{borelfeature}
\hat{B}^{(\omega)}_T \frac{1}{s-\omega}=e^{-s/T}, \hspace{2cm}
\hat{B}^{(\omega)}_{T} e^{\lambda \omega}=\delta(\lambda-\frac{1}{T}),
\end{eqnarray}
one gets
\begin{eqnarray}
\label{sumall}
2iF(A_1 v^\mu + B_1 \hat{p}^\mu)e^{-2\bar{\Lambda}_B/T}
=\int^{s_0}_{0} ds
  e^{-s/T}
\Big[\rho_a(\xi ,s)v^\mu+\rho_b(\xi,s) \hat{p}^\mu \Big]
\end{eqnarray}
with
\begin{equation}
\label{Borel2kernal}
\rho_a(\xi,s)v^\mu+\rho_b(\xi,s)\hat{p}^\mu
=\hat{B}^{(-1/T)}_{1/s} \hat{B}^{(\omega)}_{T}F^\mu(\xi,\omega)  .
\end{equation}

The operator $1/(iv\cdot D)$ in Eq.(\ref{correlator})
may be effectively regarded as a heavy quark
propagator \cite{wwy,excit}. Namely, in the following we conveniently
calculate the correlator (\ref{correlator}) via the three-point
one:
\begin{eqnarray}
\label{3pointcorrelator}
&& i\int d^4x
e^{i(q-m_Q v)\cdot x}\langle \pi(p)|T\Big\{\bar{u}(x)\gamma^\mu \QV(x),
 i^3 \int d^4y
\QVB(y) \Big(D^2_\bot +\frac{i}{2} \sigma_{\alpha\beta}
 F^{\alpha\beta} \Big)\QV(y), \nonumber\\
&&\hspace{2cm} \QVB(0)i\gamma^5 d(0) \Big\}|0 \rangle  .
\end{eqnarray}
In calculating the three-point function (\ref{3pointcorrelator}),
we represent the nonperturbative contributions embeded in the hadronic
matrix element in terms of light cone wave functions up to twist 4.
Among them are the two-particle distribution functions defined by
\cite{var,ar,vvar}
\begin{eqnarray}
\label{dadef1}
&&\langle \pi(p)|\bar{u}(x)\gamma^\mu \gamma^5 d(0)|0 \rangle
=-ip^\mu f_\pi \int^1_0 du e^{iup\cdot x}
   [\varphi_\pi(u)+x^2 g_1(u) ]\nonumber\\
&&\hspace{6cm}  +f_\pi (x^\mu-\frac{x^2 p^\mu}{x\cdot p})
   \int^1_0 du e^{iup\cdot x} g_2(u), \nonumber\\
&&\langle \pi(p)|\bar{u}(x)i \gamma^5 d(0)|0 \rangle
 =\frac{f_\pi m^2_\pi}{m_u+m_d} \int^1_0 du e^{iup\cdot x} \varphi_p(u), \nonumber\\
&&\langle \pi(p)|\bar{u}(x) \sigma_{\mu\nu} \gamma^5 d(0)|0 \rangle
= i(p_\mu x_\nu-p_\nu x_\mu)
   \frac{f_\pi m^2_\pi}{6 (m_u+m_d) } \int^1_0 du e^{iup\cdot x} \varphi_\sigma(u),
\end{eqnarray}
where $\varphi_\pi$ is the leading twist 2 distribution amplitude. $\varphi_p$ and
$\varphi_\sigma$ are twist 3 distribution amplitudes, while $g_1$ and $g_2$ are of twist 4.
Furthermore, we would also include
three-particle distribution functions \cite{vvar,bf,pb},
\newcommand{\DALPHA}{{\cal D}\alpha_i }
\begin{eqnarray}
\label{dadef2}
&&\langle \pi(p)|\bar{u}(x)\sigma_{\mu\nu}\gamma^5
   F_{\alpha\beta}(\omega x)d(0)|0\rangle =-f_{3\pi}
  [(p_\alpha p_\mu g_{\beta\nu}-p_\beta p_\mu g_{\alpha\nu})
   -(p_\alpha p_\nu g_{\beta\mu} \nonumber\\
&&\hspace{2cm} -p_\beta p_\nu g_{\alpha\mu})  ]
  \int \DALPHA \varphi_{3\pi}(\alpha_i)e^{i(\alpha_1+\omega \alpha_3)
  p\cdot x }, \nonumber\\
&&\langle \pi(p)|\bar{u}(x)\gamma_\mu \gamma^5 F_{\alpha\beta}(\omega x)
  d(0)|0\rangle =if_\pi [p_\beta (g_{\alpha\mu}-\frac{x_\alpha p_\mu}
  {p\cdot x})-p_\alpha (g_{\beta\mu}-\frac{x_\beta p_\mu}{p\cdot x})]
  \nonumber\\
&&\hspace{2cm}
\times  \int \DALPHA \varphi_\bot (\alpha_i)e^{i(\alpha_1+\omega \alpha_3)p\cdot x}
  +if_\pi \frac{p_\mu}{p\cdot x}(p_\alpha x_\beta-p_\beta x_\alpha)\nonumber\\
&&\hspace{2cm}\times  \int \DALPHA \varphi_{\|}(\alpha_i)e^{i(\alpha_1+\omega \alpha_3)p\cdot x}, \nonumber\\
&&\langle \pi(p)|\bar{u}(x)\gamma_\mu \tilde{F}_{\alpha\beta}(\omega x)
   d(0)|0\rangle =-f_\pi [p_\beta (g_{\alpha\mu}-\frac{x_\alpha p_\mu}
   {p\cdot x})-p_\alpha (g_{\beta\mu}-\frac{x_\beta p_\mu}{p\cdot x})]
  \nonumber\\
&&\hspace{2cm} \times  \int \DALPHA \tilde{\varphi}_\bot(\alpha_i)e^{i(\alpha_1+\omega \alpha_3) p\cdot x}  -f_\pi \frac{p_\mu}{p\cdot x}(p_\alpha x_\beta-p_\beta
   x_\alpha) \nonumber\\
&&\hspace{2cm} \times \int \DALPHA \tilde{\varphi}_{\|}(\alpha_i) e^{i(\alpha_1
   +\omega \alpha_3)p\cdot x} ,
\end{eqnarray}
where $
\DALPHA =d\alpha_1 d\alpha_2 d \alpha_3
      \delta(1-\alpha_1-\alpha_2-\alpha_3) $ and
$\tilde{F}_{\alpha\beta}=\frac{1}{2}\epsilon_{\alpha\beta\rho\sigma}
F^{\rho\sigma}$.
$\varphi_{3\pi}$ is a distribution amplitude of
twist 3, while $\varphi_{\bot}$, $\varphi_{\|}$,
$\tilde{\varphi}_{\bot}$ and
$\tilde{\varphi}_{\|}$ are of twist 4.

After contracting the effective heavy quark fields
$\QV(x_1)\QVB(x_2)$ into propagator $P_+ \int^{\infty}_{0}dt
\delta(x_1-x_2-vt)$, the correlation function
(\ref{3pointcorrelator}) becomes
\begin{eqnarray}
\label{midcal}
&& i\int d^4x \int d^4y \int^{\infty}_0 d l \int^{\infty}_0 dt
e^{i(q-m_Q v)\cdot x} \langle \pi(p)|\bar{u}(x)\gamma^\mu
P_+ \delta(x-y-vl) \nonumber\\
&& \times  \Big[\partial^2_{(y)}
-v_\alpha v_\beta \partial^\alpha_{(y)} \partial^\beta_{(y)}
-\partial^\alpha_{(y)} A_\alpha (y)
-A_\alpha (y) \partial^\alpha_{(y)}
+v_\alpha v_\beta \partial^\alpha_{(y)} A^\beta (y) \nonumber\\
&&+v_\alpha v_\beta A^\alpha (y) \partial^\beta_{(y)}
 +\frac{i}{2}
\sigma_{\alpha \beta} F^{\alpha \beta} (y) \Big]
  P_+ \delta(y-vt) \gamma^5 d(0)|0 \rangle
\end{eqnarray}
with $\partial^\alpha_{(y)} \equiv \partial /
(\partial y_\alpha)$,
which includes only terms related to the two-particle and
three-particle
distribution functions (\ref{dadef1}) and (\ref{dadef2}),
while other terms are not considered in this paper.

Then one can represent the matrix elements by the distribution
functions defined in Eqs.(\ref{dadef1}) and (\ref{dadef2}).
In doing this we choose to work in the fixed-point gauge
$x^\mu A_\mu(x)=0$, where one has
\[ A_\mu(x)=x^\nu \int^1_0 d\omega \omega F_{\nu\mu}(\omega x) . \]
Using Eqs. (\ref{borelfeature}), (\ref{dadef1}) and (\ref{dadef2}),
we finally get
\begin{eqnarray}
\label{srrhoa}
\rho_a(\xi,s)&=&\frac{i}{12 \xi^3}f_\pi \Big[
  36 \mu_\pi (u-1) \xi \tilde{\xi}^2 \varphi_p +18 \mu_\pi (u-1)^2
  \xi \tilde{\xi}^2 \varphi'_p
  -6 \mu_\pi m^2_\pi \xi \varphi_\sigma \nonumber\\
&&  +\mu_\pi (u-1)\xi (10 \tilde{\xi}^2-2m^2_\pi)\varphi'_\sigma
  +3\mu_\pi (u-1)^2 \xi \tilde{\xi}^2\varphi''_\sigma +36 m^2_\pi g_2
   \nonumber\\
&&  -(u-1)(60 \tilde{\xi}^2 -12 m^2_\pi) g'_2 -18 (u-1)^2 \tilde{\xi}^2
  g''_2   \Big]_{u=1-\frac{s}{2\xi}} \nonumber\\
&&+\frac{i}{2\xi^3}f_\pi \int^1_0 d\alpha_3 \frac{1}{\alpha_3}
 \Big [ \Big[2\xi^2(\tilde{\varphi}_\bot-\varphi_\bot)+m^2_\pi (\tilde{\varphi}_\bot
   +\tilde{\varphi}_{\|})\Big]_{\alpha_2=1-\alpha_1-\alpha_3}
 \Big ]^{\alpha_1=1-\alpha_3-\frac{s}{2\xi}}_{\alpha_1=1-\frac{s}{2\xi}} ,
 \\
\label{srrhob}
\rho_b(\xi,s)&=&  \frac{i}{12 \xi^3} f_\pi
\Big[ 36 (u-1) \xi^2 \tilde{\xi}^2 \varphi_\pi
 +18 (u-1)^2 \xi^2 \tilde{\xi}^2 \varphi'_\pi +2 \mu_\pi \xi (\xi^2+3 m^2_\pi)
 \varphi_\sigma  \nonumber\\
&& -\mu_\pi (u-1)\xi (10 \tilde{\xi}^2-2m^2_\pi) \varphi'_\sigma
 -3\mu_\pi (u-1)^2 \xi \tilde{\xi}^2 \varphi''_\sigma +36(\xi^2+3m^2_\pi)g'_1
  \nonumber\\
&&  -108(u-1)\tilde{\xi}^2 g''_1 -18(u-1)^2 \tilde{\xi}^2g'''_1
 -12(2\xi^2+5m^2_\pi)g_2  \nonumber\\
&& +(u-1)(72 \tilde{\xi}^2-12 m^2_\pi)g'_2
  +18(u-1)^2\tilde{\xi}^2 g''_2 \Big ]_{u=1-\frac{s}{2\xi}} \nonumber \\
&&  -\frac{i}{\xi^3} f_\pi \int^1_0 d\alpha_3 \int^1_0 du \int^1_0 d\rho
   \frac{1}{\alpha_3 } (2\xi^2+m^2_\pi)
 \Big [\varphi_\bot+\varphi_{\|} \Big]_{
  \alpha_1=1-u\rho \alpha_3-\frac{s}{2\xi},
\alpha_2=(u\rho-1)\alpha_3 +\frac{s}{2\xi}
} \nonumber \\
&& +\frac{i}{2\xi}f_\pi \int^1_0 d\alpha_3 \frac{1}{\alpha_3}
  \Big[ \Big[2\varphi_\bot -\tilde{\varphi}_{\|}\Big]_{\alpha_2=1-\alpha_1-\alpha_3}
\Big]^{\alpha_1=1-\alpha_3-\frac{s}{2\xi}}_{\alpha_1=1-\frac{s}{2\xi}} ,
\end{eqnarray}
where $\tilde{\xi}^2 \equiv \xi^2-m^2_\pi$, and
for arbitrary function $f(x,y)$ we use the symbols
$[f(x,y)]_{x=x_0} \equiv f(x_0)$ and $[f(x,y)]^{x=x_1}_{x=x_2}
\equiv f(x_1,y)-f(x_2,y)$.
For the detailed techniques used in evaluating
Eqs.(\ref{srrhoa})-(\ref{srrhob}), we refer to Ref.\cite{bpi}.

\section{Numerical analysis}\label{numericalana}

With the scale dependence written explicitly, the distribution
amplitudes read \cite{ar,vvar,bf,pb,vi}
\begin{eqnarray}
\label{pifunction}
\varphi_\pi(u,\mu)&=& 6u(1-u)\Big\{1+\frac{3}{2}a_2(\mu)[5(2u-1)^2-1]+\frac{15}{8}
   a_4(\mu)
  [21 (2u-1)^4 \nonumber\\
&& -14 (2u-1)^2+1]\Big\} , \nonumber\\
\varphi_p(u,\mu)&=& 1+\frac{1}{2}B_2(\mu) [3(2u-1)^2-1]+\frac{1}{8}B_4(\mu) [35 (2u-1)^4
  -30 (2u-1)^2+3] ,\nonumber\\
\varphi_\sigma(u,\mu)&=& 6u(1-u) \Big\{ 1+\frac{3}{2}C_2(\mu) [5(2u-1)^2-1]+\frac{15}{8}C_4(\mu)
  [21 (2u-1)^4 \nonumber\\
&&-14(2u-1)^2+1] \Big\},  \nonumber\\
g_1(u,\mu)&=& \frac{5}{2}\delta^2(\mu) u^2 (1-u)^2+\frac{1}{2}\varepsilon(\mu) \delta^2(\mu)
  [u(1-u)(2+13u (1-u) ) \nonumber\\
&& +10u^3 \log u(2 -3u+\frac{6}{5}u^2)
  +10(1-u)^3 \log((1-u)(2-3(1-u) \nonumber\\
&& +\frac{6}{5}(1-u)^2)) ], \nonumber\\
g_2(u,\mu)&=&\frac{10}{3} \delta^2(\mu) u(1-u)(2u-1), \nonumber\\
\varphi_{3\pi}(\alpha_i,\mu)&=&360 \alpha_1 \alpha_2 \alpha^2_3 [ 1+
   \frac{1}{2}\omega_{1,0}(\mu)(7\alpha_3-3)+\omega_{2,0}(\mu)(2-4\alpha_1
    \alpha_2-8\alpha_3+8 \alpha^2_3) \nonumber\\
 && + \omega_{1,1}(\mu)(3\alpha_1 \alpha_2
   -2\alpha_3+3\alpha^2_3)], \nonumber\\
\varphi_{\bot}(\alpha_i,\mu)&=&30 \delta^2(\mu)(\alpha_1-\alpha_2)\alpha^2_3
  [\frac{1}{3}+2\varepsilon(\mu)(1-2\alpha_3) ], \nonumber\\
\varphi_{\|}(\alpha_i,\mu)&=&120 \delta^2(\mu)\varepsilon(\mu)(\alpha_1-\alpha_2)
   \alpha_1 \alpha_2 \alpha_3, \nonumber\\
\tilde{\varphi}_{\bot}(\alpha_i,\mu)&=&30 \delta^2(\mu)\alpha^2_3(1-\alpha_3)
   [\frac{1}{3}+2\varepsilon(\mu)(1-2\alpha_3)], \nonumber\\
\tilde{\varphi}_{\|}(\alpha_i,\mu)&=&-120 \delta^2(\mu) \alpha_1 \alpha_2 \alpha_3
   [\frac{1}{3}+\varepsilon(\mu)(1-3\alpha_3) ] .
\end{eqnarray}
The asymptotic form of these functions and the renormalization scale
dependence are given by perturbative QCD \cite{bf,va}.
In the current study, $\mu$ should be a typical scale of
$B\to \pi l\nu$ decay. We set $\mu$ to the typical virtuality of the
bottom quark,
\begin{equation}
\label{mub}
\mu_b=\sqrt{m^2_B-m^2_b} \approx 2.4\mbox{GeV} ,
\end{equation}
which is the same as that used in Refs.\cite{ar,vvar}.
Accordingly, the parameters appearing in the distribution amplitudes
(\ref{pifunction}) are taken as \cite{ar,vvar}
\begin{eqnarray}
\label{para}
a_2(\mu_b)=0.35, \;\; a_4(\mu_b)=0.18, \;\; B_2(\mu_b)=0.29, \;\;
B_4(\mu_b)=0.58, \;\;\; \;  \nonumber\\
C_2(\mu_b)=0.059, \;\; C_4(\mu_b)=0.034, \;\;
\delta^2(\mu_b)=0.17 \mbox{GeV}^2, \;\; \varepsilon(\mu_b)=0.36 .
\end{eqnarray}
Other meson quantities needed in the sum rule study have been
studied via sum rules and other
approaches. Here we use $\mu_\pi=2.02$GeV,
$f_\pi=0.132$GeV\cite{ar,vvar},
$\bar\Lambda_B=0.53$GeV and $F=0.30\mbox{GeV}^{3/2}$ \cite{ww}.

With these quantities,
the variation of $1/m_Q$ order wave functions with respect
to the Borel parameter $T$ is shown in Fig.1, where $v\cdot p=2.64$
GeV is fixed. Appropriate values of the threshold $s_0$
should be determined by the stability of the wave functions. Due to the
general criterion that both the higher nonperturbative corrections
and the contributions from excited and continuum states should not exceed
30\%, we focus on the region around $T=1.5$GeV. In this region, as
shown in Fig.1, there is no corresponding value of threshold which
makes the curves of wave funtions very flat. 
We choose $s_0 \approx 1.8$GeV and $s_0\approx 0.8 $GeV respectively,
where the wave functions $A_1$ and $B_1$ become relatively stable,
though the stability here is not as optimistic as that for the
leading order wave functions $A$ and $B$\cite{bpi}.

Using the thresholds given above, the wave functions are obtained
straightforwardly from Eqs.(\ref{sumall}),
(\ref{srrhoa}) and (\ref{srrhob}).
And the form factors $f_+(q^2)$ and $f_0(q^2)$
can be easily evaluated from the relations (\ref{f0ff}) and (\ref{AandL1}).
However, it is known that the light cone expansion and the
sum rule method would break down as $q^2$ approaches near
$m^2_b$\cite{ar}. At large $q^2$ (or very small $y$),
the results directly evaluated from the sum rule
(\ref{sumall}) turn
out to be not trustworthy.
In order to probe the dynamics of the whole kinematically allowed
region, we have to extrapolate the sum rule results in the
small $q^2$ region to large $q^2$ region in an appropriate way.
For this purpose, for $f_+(q^2)$ we use the sum rule results at small momentum
transfer, whereas at large $q^2$ we assume the single pole
approximation \cite{ar}
\begin{equation}
\label{sinpole}
f_+(q^2)=\frac{f_{B^*} g_{B^*B\pi}}{2m_{B^*}(1-q^2/m^2_{B^*})  } ,
\end{equation}
where $m_{B^*}=5.325$GeV, $f_{B^*} \sim 0.16$GeV
and $g_{B^*B\pi} \sim 29$\cite{ar} would be used.
Furthermore, it is convenient to parametrize the form factors as
\begin{eqnarray}
\label{fitform}
F(q^2)=\frac{F(0)}{1-a_F q^2/m^2_B+b_F q^4/m^4_B},
\end{eqnarray}
where $F$ can be $f_+$ and $f_0$, respectively. 
We fit the parameters $a_{f_+}$
and $b_{f_+}$ by using the sum rules and Eq.(\ref{sinpole}),
i.e., requiring Eq.(\ref{fitform})
approach the sum rule results at $q^2<15\mbox{GeV}^2$ but be
compatible with Eq.(\ref{sinpole}) at $q^2>15\mbox{GeV}^2$.
$f_0$ is not important in extracting $|V_{ub}|$, because
when the lepton masses are neglected, the $B\to \pi l\nu$ decay
width depends only on $f_+$. So we use
the low $q^2$ ($q^2 < 15 \mbox{GeV}^2$) region results from
LCSR to fix the relevant parameters for $f_0$ in Eq.(\ref{fitform}).
Since $f_0$ is nearly constant in $q^2$ \cite{pcps},
 one may expect this parameterization yield reasonable approximation
for $f_0$ in the whole kinematical region.

In this way we get the fitted parameters in table 1.
Then the variation of $f_+$ and $f_0$ with respect to the momentum
transfer $q^2$ are plotted in Fig.2.
In table 2 we present values of wave functions at some
kinematical points of momentum transfer.


\begin{center}
\begin{tabular}{c|c|c|c|c}
\hline \hline
 & \hspace{0.9cm} $F(0)$ \hspace{0.9cm} & \hspace{0.9cm} $a_F$ \hspace{0.9cm} &
\hspace{0.9cm} $b_F$ \hspace{0.9cm} & \hspace{0.5cm}  \hspace{0.9cm} \\
\hline
$f_+$ & $0.35$   & $1.31 $
     & $0.35$ & LO\\
\cline{2-5}
  & $0.38 $  & $1.19 $ & $0.25 $
     & NLO \\
\hline
$f_0$ & $0.35 $   & $0.60  $
     & $-0.19 $ & LO\\
\cline{2-5}
  & $0.38 $  & $0.71 $ & $0.41 $
     & NLO \\
\hline \hline
\end{tabular}
\end{center}

\vspace{0cm}
\centerline{
\parbox{13cm}{
\small
\baselineskip=1.0pt
Table 1. Results of LCSR calculations up to leading (LO) and
next leading order (NLO) in HQEFT.
The leading order results are obtained in Ref.\cite{bpi}.
} }

\vspace{0.5cm}

\begin{center}
\begin{tabular}{c|c|c|c|c|c|c}
\hline \hline
 $q^2$($\mbox{GeV}^2$) & \hspace{0.4cm}$0$ \hspace{0.4cm} & \hspace{0.4cm} $6$ \hspace{0.4cm} &
\hspace{0.4cm} $12$ \hspace{0.4cm} & \hspace{0.4cm} $18$ \hspace{0.4cm} &\hspace{0.4cm} $24$ \hspace{0.4cm}& \hspace{2cm} \\
\hline
$f_+$  & $0.38$ & $0.51$ & $0.72$   & $1.16$
     & $2.50$ & HQEFT-NLO-LCSR  \\
\cline{2-7}
    & $0.35$ & $0.48$ & $0.71$   & $1.18$
     & $2.70$ & HQEFT-LO-LCSR \cite{bpi} \\
\cline{2-7}
 & $0.42$ & $0.54$ & $0.77$   &
     &  & HQET-NLO-LCSR \cite{jcc}\\
\cline{2-7}
 & $0.36 $ & $0.47$ & $0.68$   &
     &  & HQET-LO-LCSR \cite{jcc} \\
\cline{2-7}
 & $0.28  $ &  &   &
     &  & QCD-LCSR \cite{arsc}\\
\cline{2-7}
 & $0.27 $ &  &   &
     &  & QCD-LCSR \cite{ar}\\
\cline{2-7}
 & $0.24 $ &  &   &
     &  & QCD-SR \cite{pcps}\\
\hline
$f_0$ & $0.38$ & $0.44$ & $0.50$   & $0.54$
     & $0.55$ & HQEFT-NLO-LCSR\\
\cline{2-7}
 & $0.35$ & $0.41$ & $0.50$   & $0.66$
     & $1.02$ & HQEFT-LO-LCSR \cite{h2l} \\
\hline \hline
\end{tabular}
\end{center}

\vspace{0cm}
\centerline{
\parbox{13cm}{
\small
\baselineskip=1.0pt
Table 2. Values of the form factors $f_+$ and $f_0$ at some
kinematical points.
} }

\vspace{0.5cm}

Neglecting lepton mass, the differential decay width of
$B\to \pi l\nu$ is given by
 \begin{eqnarray}
\label{gammaq2}
\frac{d\Gamma}{dq^2}=\frac{G^2_F |V_{ub}|^2}{24 \pi^3} (E^2_\pi-m^2_\pi)^{3/2}
   [f_+(q^2)]^2
\end{eqnarray}
with $E_\pi=(m^2_B+m^2_\pi-q^2)/(2m_B)$.
Now using the branching ratio
$\mbox{Br}(B^0 \to \pi^- l^+ \nu_l)=(1.8\pm 0.6)
\times 10^{-4}$
and the lifetime
$\tau_{\tiny{\mbox{B}^0}}=1.542\pm 0.016 \; ps$ \cite{pdg},
one gets from Eq.(\ref{gammaq2})
\begin{eqnarray}
\label{vubnlo}
|V_{ub}|=(3.2\pm 0.5 \pm 0.2 )\times 10^{-3},
\end{eqnarray}
where the first (second) error corresponds to the experimental (theoretical) uncertainty. Here the theoretical uncertainty is mainly
 considered from the threshold effects.
Besides this, theoretical uncertainty may arise also from using the
single pole approximation to fit the form factors at small recoil.
To take this into account effectively, if we let the couplings in
Eq.(\ref{sinpole}) change in the regions $f_{B^*}=0.16 \pm 0.03$GeV and
$g_{B^* B\pi}=29 \pm 3$ \cite{ar}, we get a more conservative value for
$|V_{ub}|$:
\begin{equation}
\label{vubnlocons}
|V_{ub}|=(3.2\pm 0.5 \pm 0.4)\times 10^{-3}.
\end{equation}
Eq.(\ref{vubnlo}) and (\ref{vubnlocons}) can be compared with the average
obtained by CLEO \cite{pdg},
\begin{equation}
|V_{ub}|=(3.25^{+0.25}_{-0.32}\pm 0.55) \times 10^{-3},
\end{equation}
where the uncertainties are statistical and experimental ones.

It is also interesting to compare our results with the relations
predicted by the large energy effective theory (LEET) \cite{efg}:
\begin{eqnarray}
\label{leet1}
f_+(q^2)&=&\frac{m^2_B+m^2_\pi}{m^2_B+m^2_\pi-q^2} f_0(q^2).
\end{eqnarray}
In Fig.3 we see that the form factor ratio $f_+/f_0$ derived from our calculations in HQEFT is compatible with that predicted by the
LEET relation (\ref{leet1}). They agree well with each other at regions not very far from the maximum recoil point.
 As a comparison, both numerical results of the 1/m order
contribution in this manuscript and that in Ref.\cite{jcc} are not large, but the former turns out to be a little
smaller than the latter. This may be attributed to the fact that we have considered the antiquark contributions and
also the input values for the involved parameters are slightly different.

\section{summary}\label{summary}

 We have studied the $1/m_Q$ order corrections
to $B\to \pi l\nu$ decay in HQEFT, while most formulae and discussions here can also be generally applied to other
heavy-to-light pseudoscalar meson decays. In HQEFT, $1/m_Q$ order corrections from the effective current and from
effective Lagrangian are given by the same operator forms. Consequently the independent wave functions are reduced
in HQEFT in comparison with the ones in HQET. These $1/m_Q$ order contributions have been calculated using light
cone sum rule with considering two-particle and three-particle distribution amplitudes up to twist 4. Numerically,
the $1/m_Q$ order wave functions give only about 10\% correction to the transition form factor $f_+$. This
correction indicates a slightly smaller value of the CKM matrix element $|V_{ub}|$. As to the form factor ratio
$f_+/f_0$, good agreement is observed between the sum rules in HQEFT and the LEET at large recoil region.

\acknowledgments

This work was supported in part by the key projects of
National Science Foundation of China (NSFC)
and Chinese Academy of Sciences,
and by the BEPC National Lab Opening Project.

\newpage
\centerline{\large{FIGURES}}

\newcommand{\PICLI}[2]
{
\begin{center}
\begin{picture}(500,120)(0,0)
\put(0,-60){
\epsfxsize=7cm
\epsfysize=10cm
\epsffile{#1} }
\put(115,10){\makebox(0,0){#2}}
\end{picture}
\end{center}
}

\newcommand{\PICRI}[2]
{
\begin{center}
\begin{picture}(300,0)(0,0)
\put(160,-30){
\epsfxsize=7cm
\epsfysize=10cm
\epsffile{#1} }
\put(275,39){\makebox(0,0){#2}}
\end{picture}
\end{center}
}


\small
\mbox{}
{\vspace{1.2cm}}

\PICLI{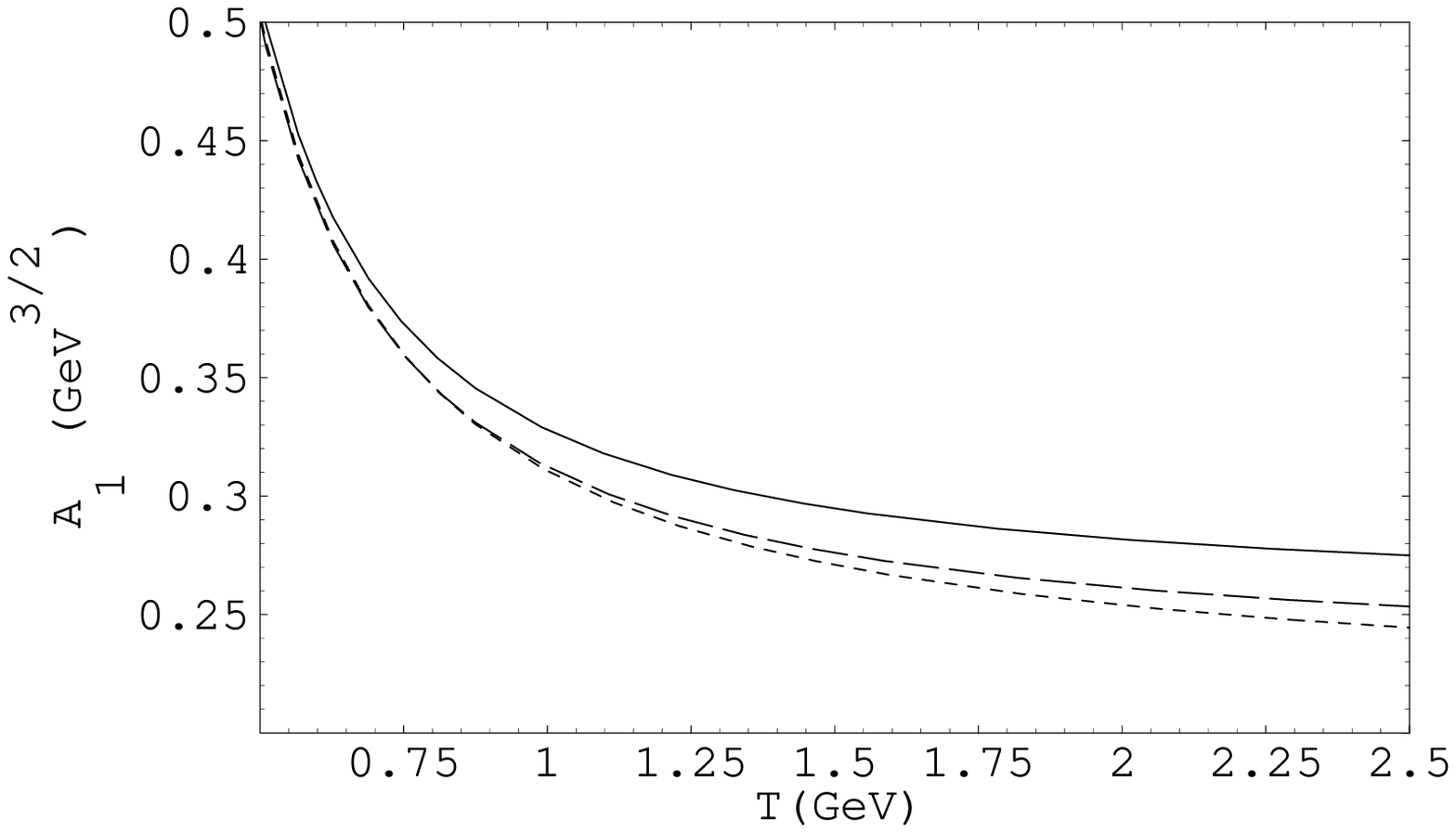}{(a)}

\PICRI{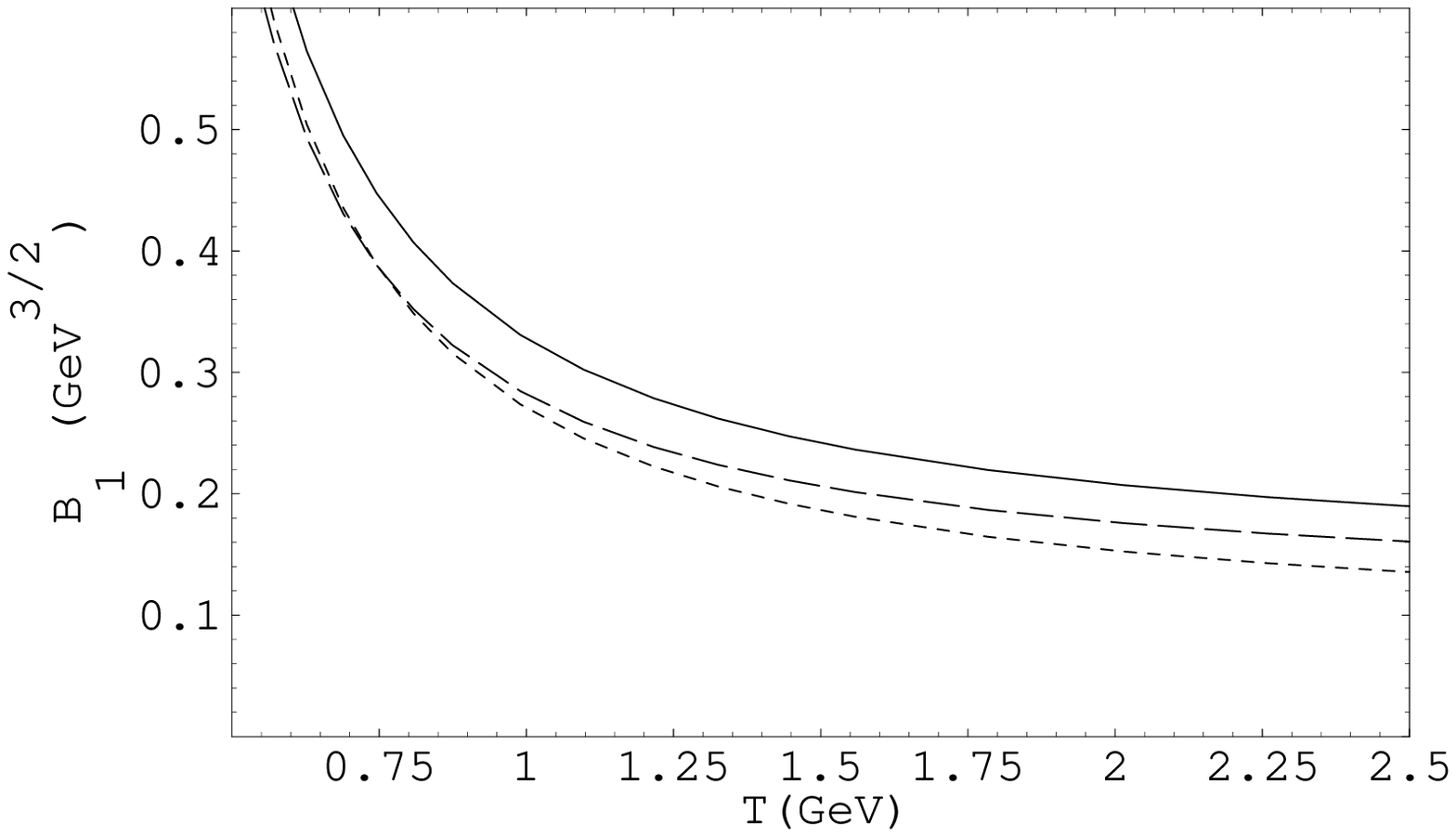}{(b)}

\vspace{-1cm}
\centerline{
\parbox{12cm}{
\small
\baselineskip=1.0pt
Fig.1. Variation of $1/m_Q$ order wave functions $A_1$ ((a)) and
$B_1$ ((b)) with respect to the Borel parameter $T$ at the momentum
transfer $q^2=0\mbox{GeV}^2$ (or $\xi=v\cdot p=2.64$GeV).
The dashed, solid and dotted curves in (a) correspond
to the thresholds $s_0=$1.5, 1.8 and 2.1 GeV respectively, while
those in (b) are derived at $s_0=$0.3, 0.8 and 1.1 GeV respectively.
}}

{\vspace{2cm}}

\PICLI{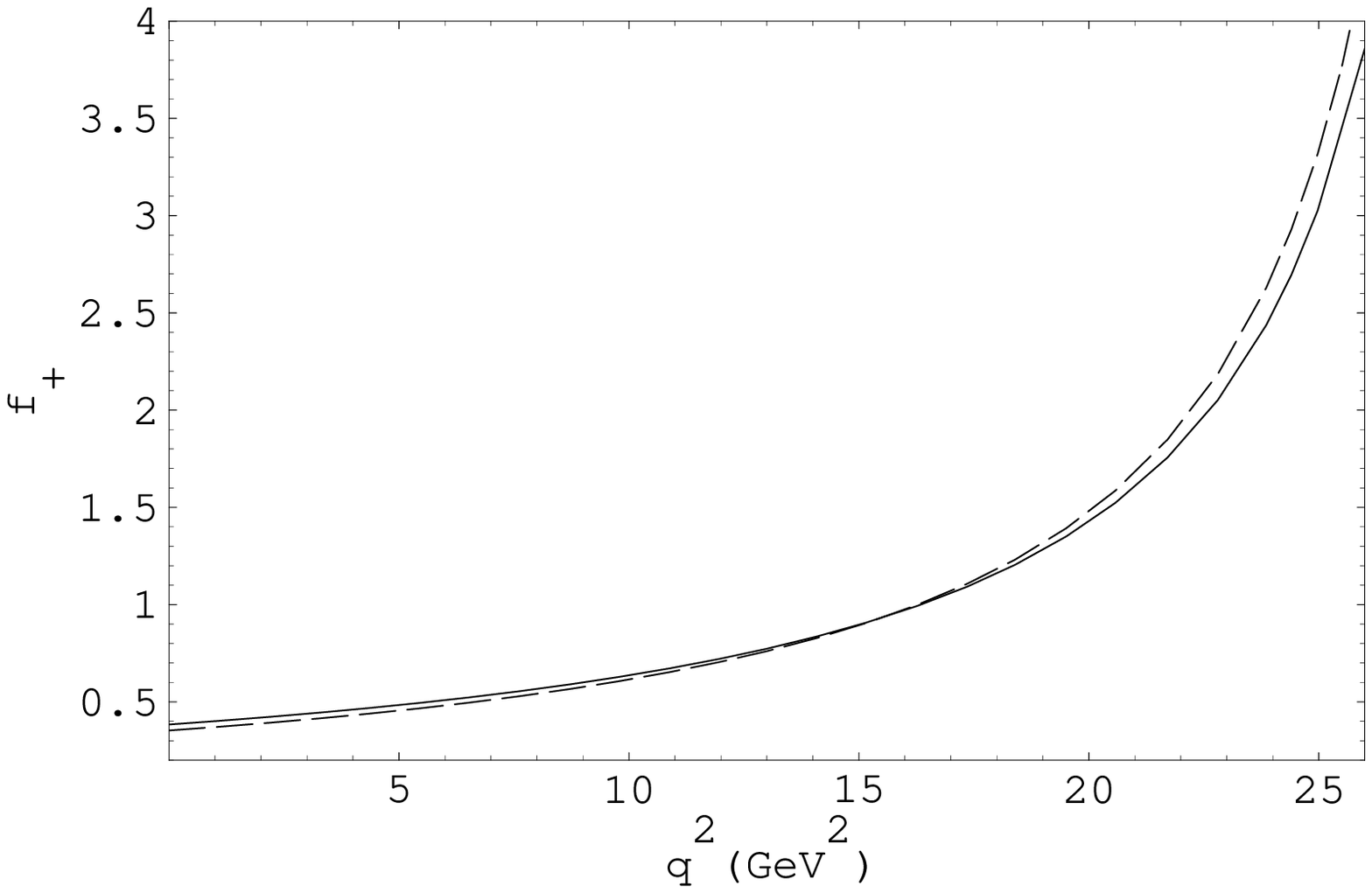}{(a)}

\PICRI{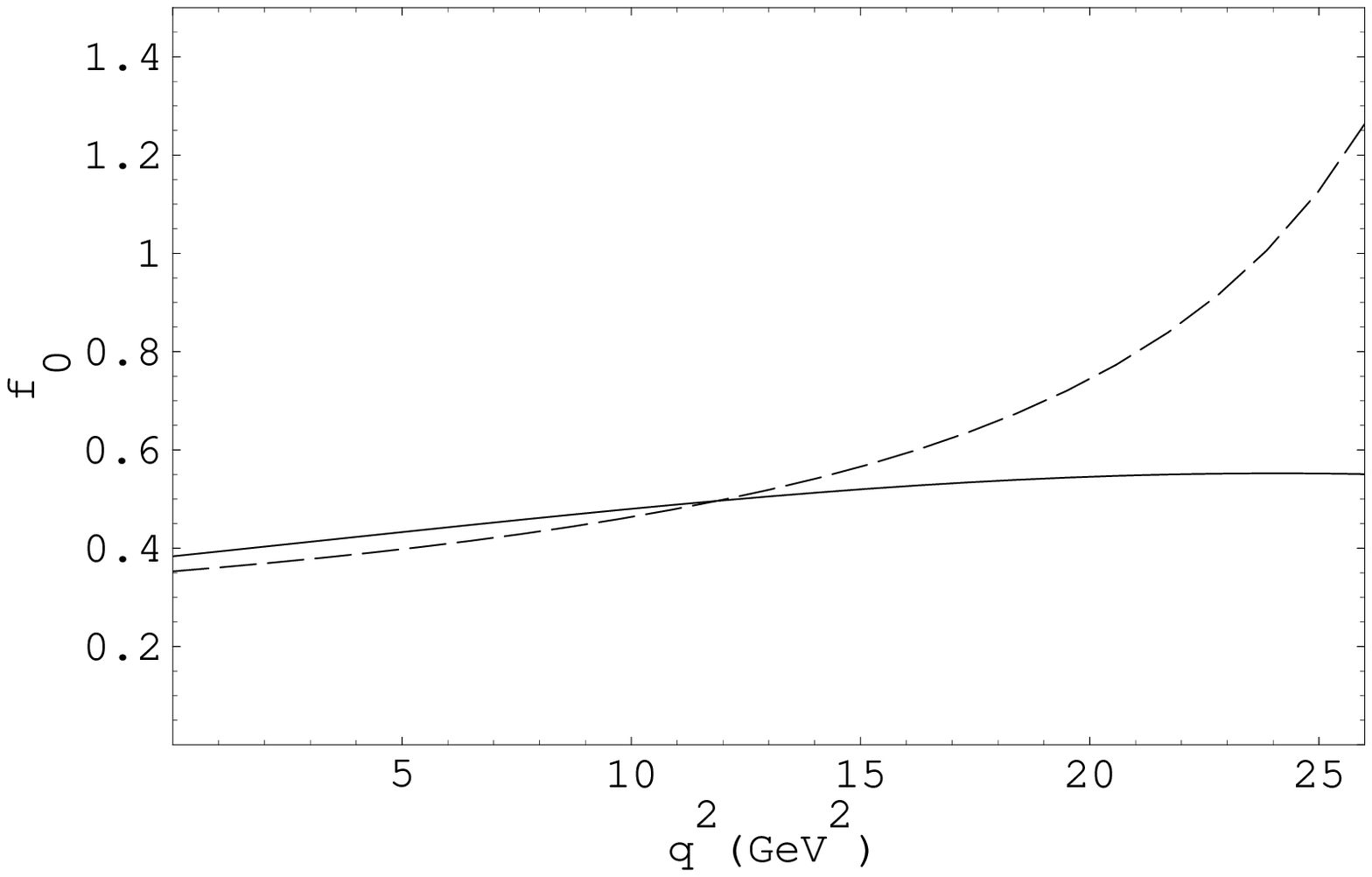}{(b)}

\vspace{-1cm}
\centerline{
\parbox{12cm}{
\small
\baselineskip=1.0pt
Fig.2. Form factors $f_+$ ((a)) and $f_0$ ((b))
obtained from light cone sum rules in HQEFT. The dashed curves
are the leading order results in HQEFT \cite{bpi,h2l}, while the solid
curves are the results with including $1/m_Q$ order correction.}}

\newpage
\mbox{}
{\vspace{1.5cm}}

\begin{center}
\begin{picture}(500,120)(0,0)
\put(125,-60){
\epsfxsize=7cm
\epsfysize=10cm
\epsffile{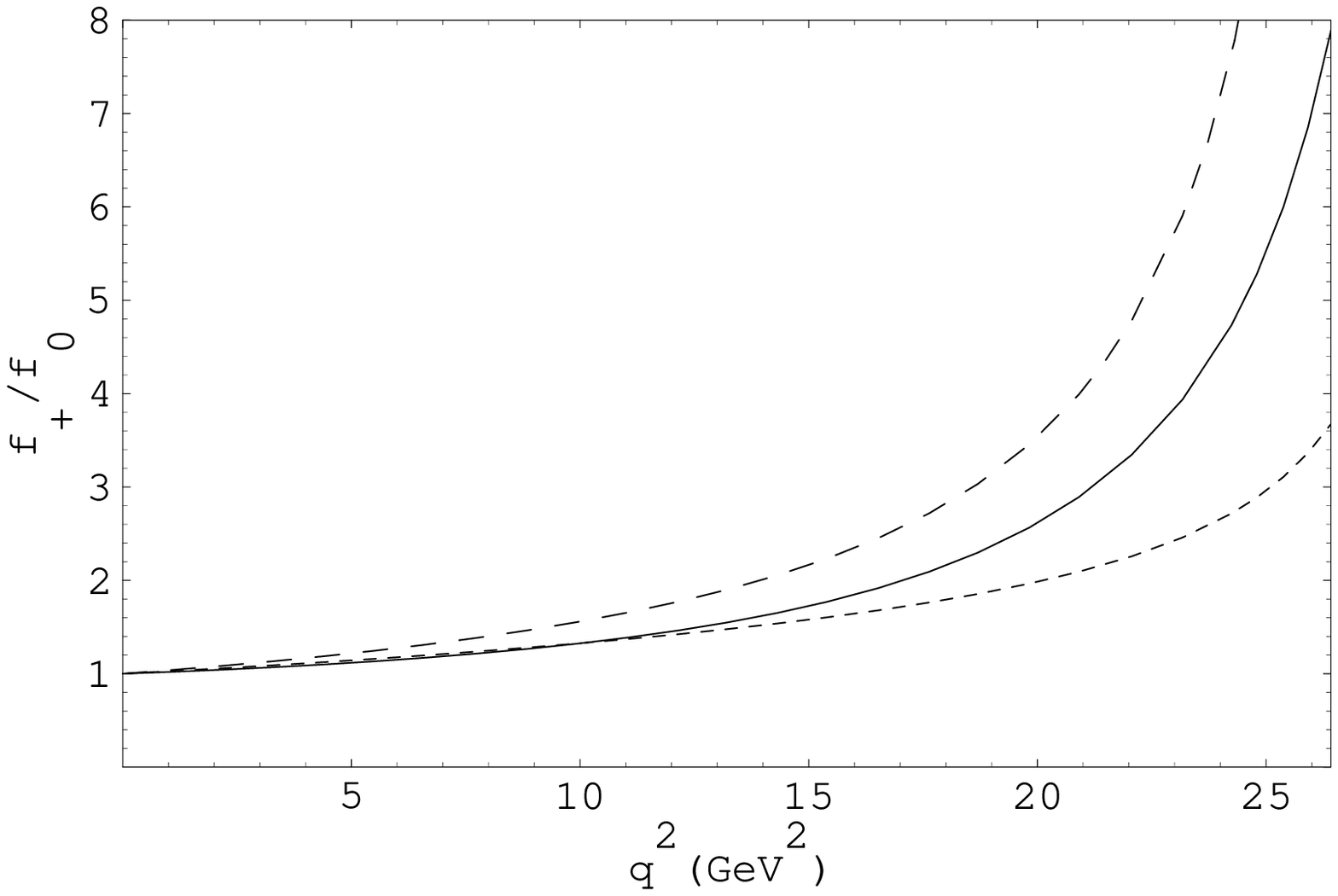} }
\put(115,10){\makebox(0,0){}}
\end{picture}
\end{center}

\vspace{-0.5cm}
\centerline{
\parbox{12cm}{
\small
\baselineskip=1.0pt
Fig.3. Comparison of the ratio $f_+(q^2)/f_0(q^2)$.
The dashed curve is obtained from the LEET relation (\ref{leet1}),
and the dotted curve is from HQEFT sum rule with including
only the leading order wave functions \cite{bpi,h2l}, while the solid
curve is the result of this paper, which includs also
 the $1/m_Q$ order contributions.}}

\end{document}